\documentclass[doublecol]{epl2}
\usepackage{graphicx}  
\usepackage{dcolumn}   
\usepackage{bm}        
\usepackage{amssymb}   
\usepackage{amsmath}
\usepackage{braket}
\usepackage[mathscr]{euscript}
\usepackage{color}
\usepackage{epsfig,color}
\usepackage{hyperref}

\title{Chaos in Qubit Coupled Optomechanical Systems}

\author{Manik Kapil \inst{1} \and Amarendra K. Sarma\inst{1}}

\institute{                   
  \inst{1} Department of Physics, Indian Institute of Technology Guwahati, Guwahati-781039, Assam, India
}

\abstract{We have found stable chaotic solutions for optomechanical systems coupled with a Two-Level System or qubit. In this system methods have been found which can be used to Tune in and out of Chaos as well as various n-period motions. This includes achieving chaos by changing the detuning, coupling parameters, and Power of the driving laser. This allows us to manipulate chaos using either the qubit or the optical cavity. Chaotic motion was also observed in both the qubit and cavity by only changing the relative phase between of driving fields of the two. This gives us the prospect of creating and exploring chaotic motion in quantum mechanical systems with further ease.}

\begin{document}

\maketitle

Cavity optomechanics has gained widespread usage in studying quantum optical and nonlinear optical phenomenon in recent years\cite{b.p1,b.p2,b.p3,b.p4,b.p5}. Optomechanics systems have gained particular interest while studying the fundamental nature of quantum mechanics\cite{b.p6,b.p7,b.p8}. For our purposes Optomechanical systems can also produce self-induced oscillations\cite{b.p9,b.p10,b.p11,b.p12,b.p13} which show both periodic and chaotic motion\cite{b.p14,b.p15,b.p16}. This chaotic motion of the optomechanical system has been studied and observed in Various systems. This includes simple passive systems\cite{b.p16,b.p17} as well as coupled optomechanical systems\cite{b.p18,b.p19} and hybrid systems\cite{b.p20,b.p21}.\\
A simple optomechanical system consists of a Fabry Perot cavity where one of the mirrors is a mechanical oscillator or a cantilever. The mechanical oscillator is affected by the radiation pressure from the optical source.  We can also construct hybrid optomechanical systems by coupling cavities with other components which could include other cavities\cite{b.p22,b.p23}, LC circuits\cite{b.p24,b.p25}, Bose-Einstein Condensates\cite{b.p26}, Two-Level Systems\cite{b.p27,b.p28}, and so on\cite{b.p29}. In this paper, we have coupled our system to a two-level system or a qubit. While this system has been studied earlier\cite{b.p27,b.p28}, we have particularly focused our research on period-doubling bifurcation and chaos.\\
Coupling an Optomechanical system to a qubit leads to various interesting phenomena. Primarily it allows us to induce chaos in either the optomechanical system or the qubit but inducing chaos in the other part. This has various uses since Chaos in both Optomechanical systems and Qubits can be used for secret communication\cite{b.p32,b.p30,b.p31}, optical sensing\cite{b.p33} or random number generation\cite{b.p34}. This could allow for using a chaotic qubit for secret communication using Quantum Computers. Furthermore, Chaos can be induced with relative ease by changing either the relative phase or the input power in the Cavity and the qubit, or by simply changing the power. This gives us various diverse ways of controlling chaos in a system using this method.\\
\begin{figure}[h!]
    \includegraphics[width=0.45\textwidth]{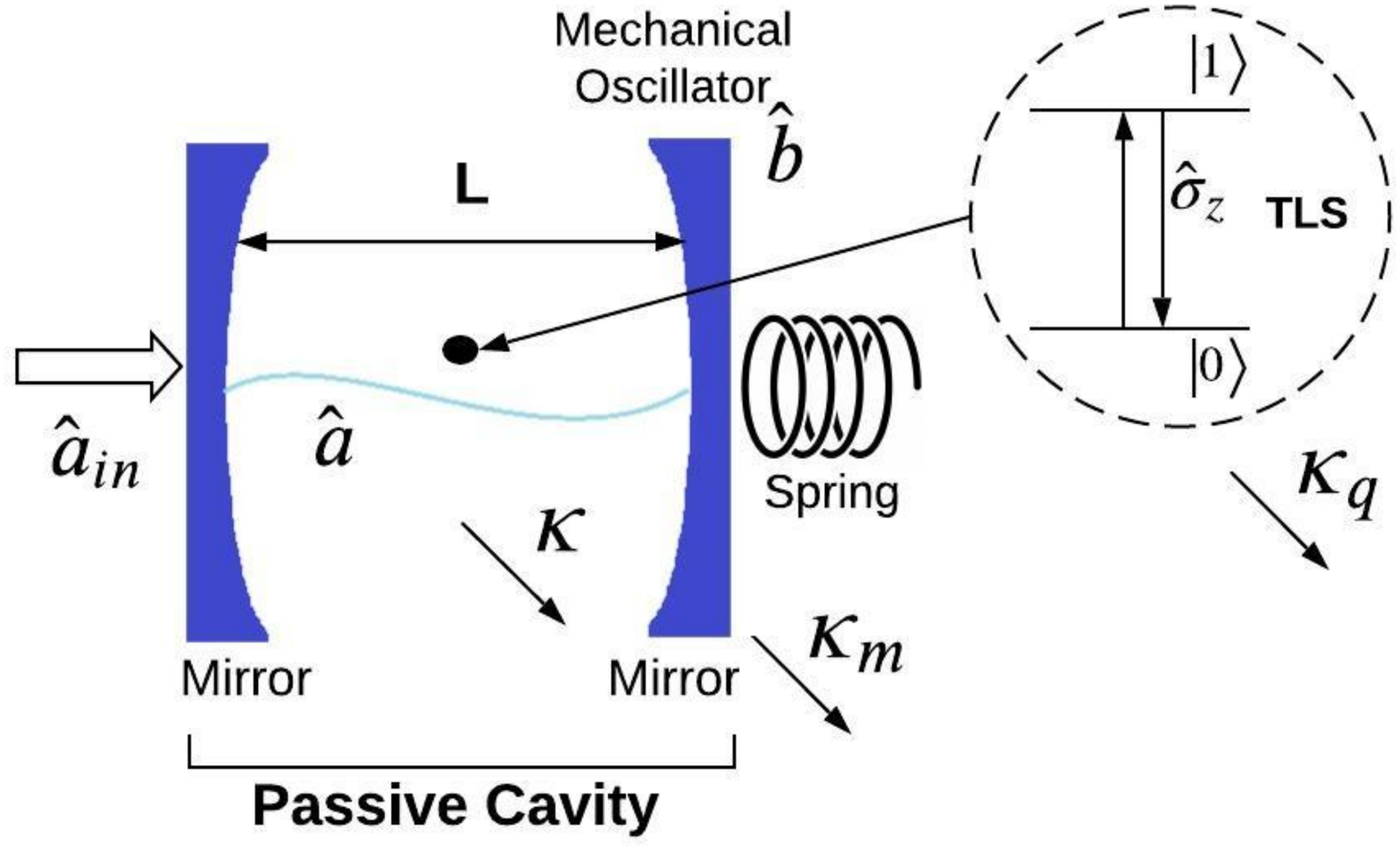}
    \caption{Optomechanical Cavity Coupled with Two Level System(Qubit). We have a cavity with two mirrors, one of which is attached to a spring and acts as a mechanical oscillator. We also have a Two Level System(TLS) coupled with the optical cavity. The decay parameters and operators have also being defined and described later.}
    \label{Fig1}
\end{figure}
In this work, we have coupled a Qubit with the optical mode as shown in \ref{Fig1}. The Hamiltonian of this system can be given as
\begin{equation}
    H=H_{p}+H_{q}+H_{coupling}
\end{equation}
where $H_{p}$ is the Hamiltonian of Passive Cavity\cite{b.p1,b.p2,b.p35}, $H_{q}$ is the Hamiltonian of Qubit\cite{b.p36,b.p37} and $H_{coupling}$ is the Hamiltonian of Coupling Parameter\cite{b.p36,b.p37}. Here the optical cavity is coupled with a mechanical oscillator and a qubit with coupling parameters $g$ and $J$ respectively. We have used $\widehat{a}(\widehat{a}^{\dagger})$, $\widehat{b}(\widehat{b}^{\dagger})$ and $\widehat{q}(\widehat{q}^{\dagger})$  represents the optical, mechanical and two level systems(Qubit) annihilation(creation) operators. We have also used the Pauli Operator to describe the state of the Qubit.\\
In this system, the resonance Frequency of the qubit is given as $\omega_{q}$. The qubit is driven by a field of Amplitude $\Omega_{q}$, and frequency $\omega_{c}$ and has a decay rate of $\kappa_{q}$. Similarly we have a cavity with a frequency of $\omega$ and is driven by a field of amplitude $\Omega$, and frequency $\omega_{p}$ and a decay rate of $\kappa$. Our mechanical oscillator also has a resonance frequency of $\omega_{m}$ and a decay rate of $\kappa_{m}$. Using the Rotating Wave Approximation we can solve the system in terms of detuning $\Delta=\omega-\omega_{p}(\Delta_{q}=\omega_{q}-\omega_{c})$ where $\omega_{p}, \omega_{c}$ represents the frame rotating frequency for the optical cavity and qubit respectively. We have also defined $\phi$ as the relative phase difference between the control field and the pumping field.
\begin{gather*}
    H_{p}=\hbar\Delta\widehat{a}^{\dagger}\widehat{a} - \hbar g\widehat{a}^{\dagger} \widehat{a}(\widehat{b}^{\dagger}+ \widehat{b})+ \hbar\omega_{m}\widehat{b}^{\dagger} \widehat{b} -i \hbar\Omega(\widehat{a}^{\dagger}-\widehat{a})\\
    H_{q}=\frac{\hbar}{2}\Delta_{q}\sigma_{z}-i \hbar\Omega_{q}(\widehat{c}^{\dagger}e^{i\phi}-\widehat{c}e^{-i\phi})\\
    H_{coupling}=\hbar J(a^{\dagger}\sigma_{-}+a\sigma_{+})
\end{gather*}
Hence our complete Hamiltonian of our system can be written as
\begin{multline}
    H=\hbar\Delta\widehat{a}^{\dagger}\widehat{a} - \hbar g\widehat{a}^{\dagger} \widehat{a}(\widehat{b}^{\dagger}+ \widehat{b})+ \hbar\omega_{m}\widehat{b}^{\dagger} \widehat{b}+\frac{\hbar}{2}\Delta_{q}\sigma_{z} \\+\hbar J(a^{\dagger}\sigma_{-}+a\sigma_{+}) -i \hbar\Omega(\widehat{a}^{\dagger}-\widehat{a})-i \hbar\Omega_{q}(\sigma_{+}e^{i\phi}-\sigma_{-}e^{-i\phi})
\end{multline}
It would be easier to simplify the system in terms of ladder operators using the Holstein-Primakoff approximation\cite{b.p37,b.p38} where $q=\sigma_{-}$. Therefore we can also write $q^{\dagger}=\sigma_{+}$. We also have $\sigma_{z}=2q^{\dagger}q-1$. Therefore we can write
\begin{multline}
    H=\hbar\Delta\widehat{a}^{\dagger}\widehat{a} - \hbar g\widehat{a}^{\dagger} \widehat{a}(\widehat{b}^{\dagger}+ \widehat{b})+ \hbar\omega_{m}\widehat{b}^{\dagger} \widehat{b}+\hbar\Delta_{q}q^{\dagger}q \\ +\hbar J(a^{\dagger}q+aq^{\dagger})-i \hbar\Omega(\widehat{a}^{\dagger}-\widehat{a})-i \hbar\Omega_{q}(\widehat{q}^{\dagger}e^{i\phi}-\widehat{q}e^{-i\phi})
\end{multline}
which can be solved using the Heisenberg-Langevin equations as
\begin{equation}
\frac{d\langle\widehat{a}\rangle}{dt} = -i\Delta\langle\widehat{a}\rangle + ig\langle(\widehat{b}^{\dagger}+ \widehat{b})\widehat{a}\rangle + iJ\langle\widehat{q}\rangle - \frac{\kappa\langle\widehat{a}\rangle}{2} +\Omega
\end{equation}
\begin{equation}
\frac{d\langle\widehat{b}\rangle}{dt} = -i\omega_{m}\langle\widehat{b}\rangle+ ig|\langle\widehat{a}\rangle|^2 - \frac{\kappa_{m}\langle\widehat{b}\rangle}{2}
\end{equation}
\begin{equation}
\frac{d\langle\widehat{q}\rangle}{dt} = -i\Delta_{q}\langle\widehat{q}\rangle + iJ\langle\widehat{a}\rangle- \frac{\kappa_{q}\langle\widehat{q}\rangle}{2}+\Omega_{q}e^{i\phi}
\end{equation}
Now we can further simplify the system by taking $\tau=\omega_{m}t$, $\alpha=\frac{\omega_{m}\langle\widehat{a}\rangle}{2\Omega}$, $\beta=\frac{g\langle\widehat{b}\rangle}{\omega_{m}}$, $\psi=\frac{\omega_{m}\langle\widehat{q}\rangle}{2\Omega}$, $P = \frac{8g^{2}\Omega^{2}}{\omega_{m}^{4}}$ and $P_{p}=\frac{\Omega_{q}}{\Omega}$\cite{b.p39,b.p40,b.p41}. Hence we get
\begin{equation}
\frac{d\alpha}{dt} = -i\frac{\Delta}{\omega_{m}}\alpha + i(\beta+\beta^{\dagger})\alpha + i\frac{J}{\omega_{m}}\psi - \frac{\kappa}{2\omega_{m}}\alpha +\frac{1}{2}
\end{equation}
\begin{equation}
\frac{d\beta}{dt} = -i\beta- i\frac{P}{2}|\alpha|^2 - \frac{\kappa_{m}}{2\omega_{m}}\beta
\end{equation}
\begin{equation}
\frac{d\psi}{dt} = -i\frac{\Delta_{q}}{\omega_{m}}\psi + i\frac{J}{\omega_{m}}\alpha- \frac{\kappa_{q}}{2\omega_{m}}\psi+\frac{P_{p}}{2}e^{i\phi}
\end{equation}
This system of equations allows us to further simplify our analysis of the system. Firstly it allows us to express our system in dimensionless parameters including P and $P_{p}$. Here the pump parameter P gives us the strength of the control field and $P_{p}$ is the ratio of the driving amplitudes of the pumping vs the control field. We shall now use this system to analyse chaos in the system. We shall further simplify our system by expressing our equations in terms of dimensionless parameters.  For future reference we shall take $\omega_{m}=1$ and consider our other parameters as dimensionless.
\begin{equation}
\frac{d\alpha}{dt} = -i\Delta\alpha + i(\beta+\beta^{\dagger})\alpha + iJ\psi - \frac{\kappa}{2}\alpha +\frac{1}{2}
\end{equation}
\begin{equation}
\frac{d\beta}{dt} = -i\beta- i\frac{P}{2}|\alpha|^2 - \frac{\kappa_{m}}{2}\beta
\end{equation}
\begin{equation}
\frac{d\psi}{dt} = -i\Delta_{q}\psi + iJ\alpha- \frac{\kappa_{q}}{2}\psi+\frac{P_{p}}{2}e^{i\phi}
\end{equation}\\
Now we can take the various values as $\frac{\Delta}{\omega_{m}}=-0.65$, $\frac{J}{\omega_{m}}=0.2$ $\frac{\Delta_{q}}{\omega_{m}}=0.5$, $\frac{\kappa}{\omega_{m}}=1$, $\frac{\kappa_{m}}{\omega_{m}}=0.001$, and $\frac{\kappa_{q}}{\omega_{m}}=1$. For future reference we shall ignore $\omega_{m}$ to keep our parameters dimensionless. We also know that $\hat{x}=\frac{\hat{\beta}+\hat{\beta^{\dagger}}}{\sqrt{2}}$ and $\hat{p}=i\frac{\hat{\beta}-\hat{\beta^{\dagger}}}{\sqrt{2}}$.
\begin{figure}[ht!]
    \includegraphics[width=0.45\textwidth]{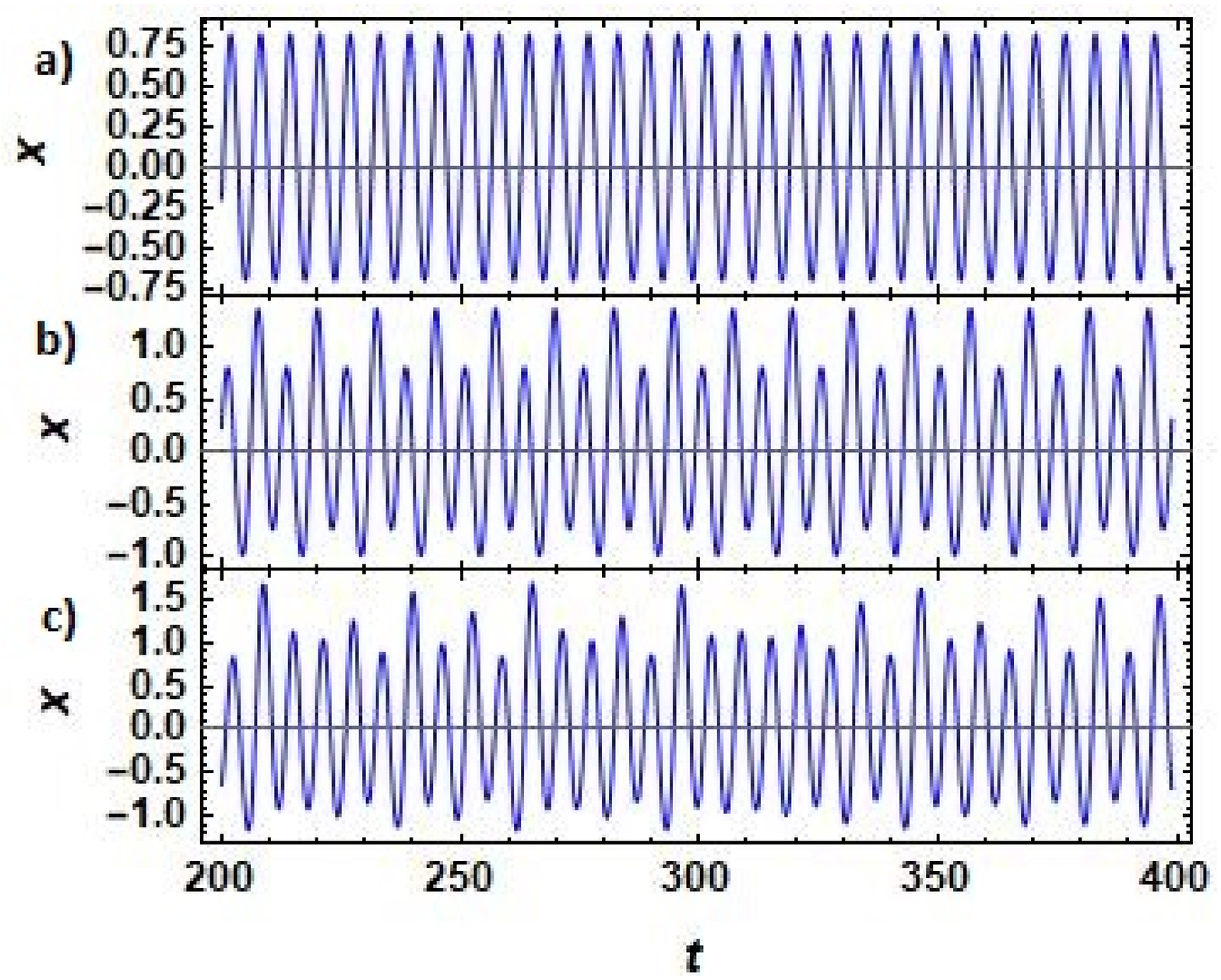}
    \caption{X vs t graph of Qubit Coupled Optomechanical System. Here we have taken P=1.4 and $P_{p}=0.5$. We also have the values of $\phi$ as a)-3, b)-2 and c)-1.3. Here we can see that for we have Period-1 motion for a), period-2 orbit for b) and chaotic motion for c). }
    \label{Fig2}
\end{figure}\\
We can see in \ref{Fig2} that as we change the value of $\phi$ it can lead to period doubling and then chaotic motion. In \ref{Fig2} we can also see that at $\phi=-3$ we observe a simple period-1 orbit. As we increase the value of $\phi$ to $\phi=-2$ it doubles the period of the motion. Continuing to increase $\phi$ we can reach n-period cycles and finally at $\phi=-1.3$ we observe chaotic motion. We can prove that the observed motion is chaotic since since we have bounded motion with a positive Maximal Lyapunov exponent(MLE)\cite{b.p42,b.p43}. This allows us to distinguish normal n-periodic motion from chaotic motion since in chaotic motion the value of MLE is positive. This also allows us to observe period doubling since at these points the value of the MLE becomes 0. We have calculated the MLE here using the standard method.
We now plot both the Lyapunov exponent and the Bifurcation diagram for the amplitude of the mechanical oscillator of these systems. The position of the mechanical oscillator can be found as $\hat{x}=\frac{\hat{\beta}+\hat{\beta^{\dagger}}}{\sqrt{2}}$. We can plot this graph for changing the values of $J_{c}$ and observe the data as given in  \ref{Fig3}
\begin{figure}[ht!]
    \includegraphics[width=0.45\textwidth]{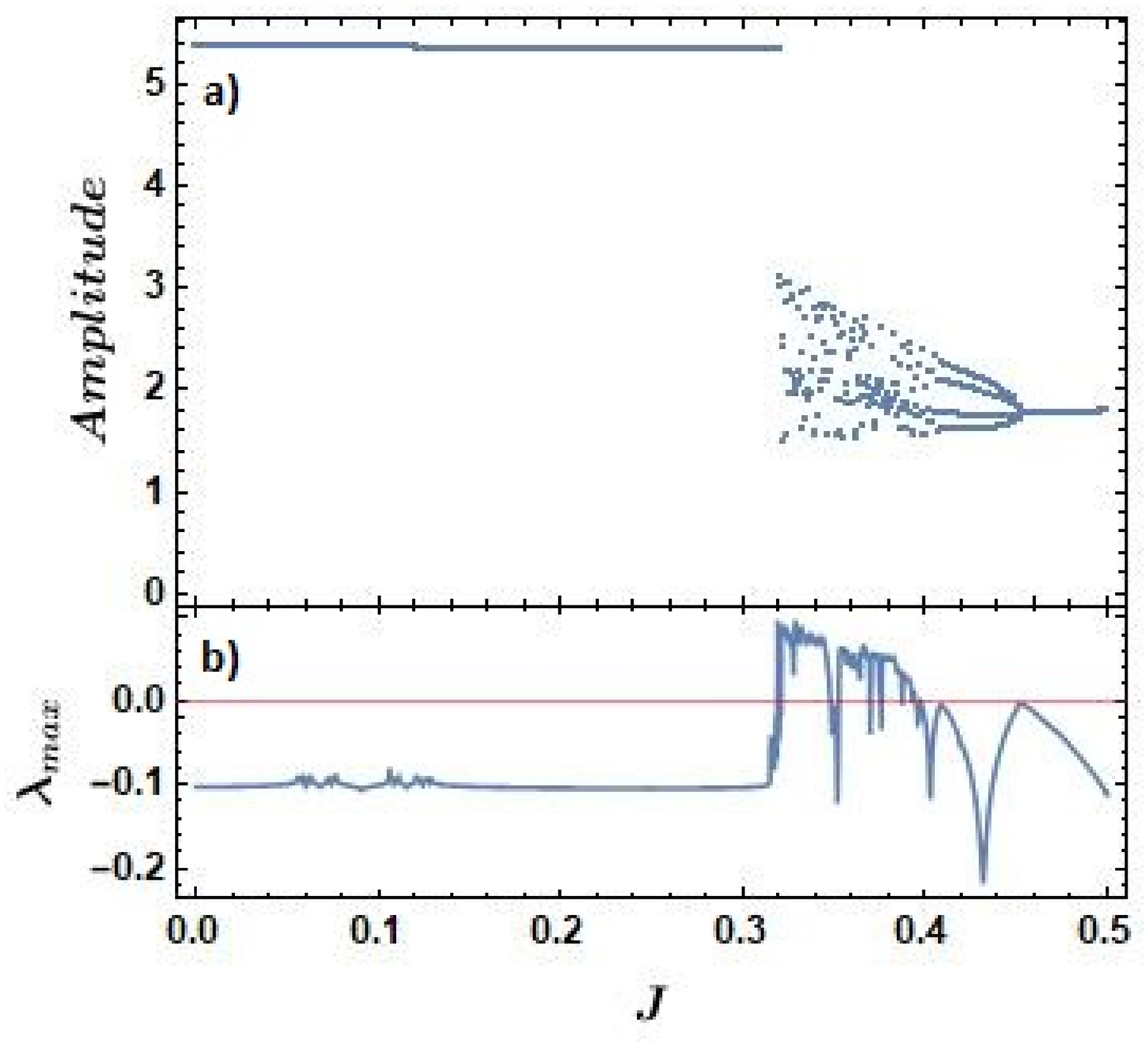}
    \caption{a) The Bifurcation Diagram for the system as we change the value of $J_{c}$ b) Maximum Lyapunov Exponent for changing values of $J_{c}$. Here $\Delta=-0.75$, $\Delta_{q}=-0.75$, $\omega_{m}=1$, $\kappa=1$, $\kappa_{m}=0.001$, $\kappa_{q}=1$, $P=2.4$, $P_{p}=0$, and $\phi=0$}
    \label{Fig3}
\end{figure}\\
In \ref{Fig3} we can see that for values less than $J=0.32$ the amplitude of the amplitude is constant. But for values larger than $J=0.32$  the motion suddenly becomes chaotic as we keep increasing $J$. After that we can again observe period-4 motion and period-1 motion. But in general changing the value of $J$ might not be practically feasible\\
This is not an issue though since we can also observe Chaotic motion by changing the detuning of the qubit ($\Delta_{q}$) as in \ref{Fig4}. Here we can again observe again chaotic motion and this time period-4 motion and period-1 motion as we change for various values of detuning.\\
\begin{figure}[ht!]
    \includegraphics[width=0.45\textwidth]{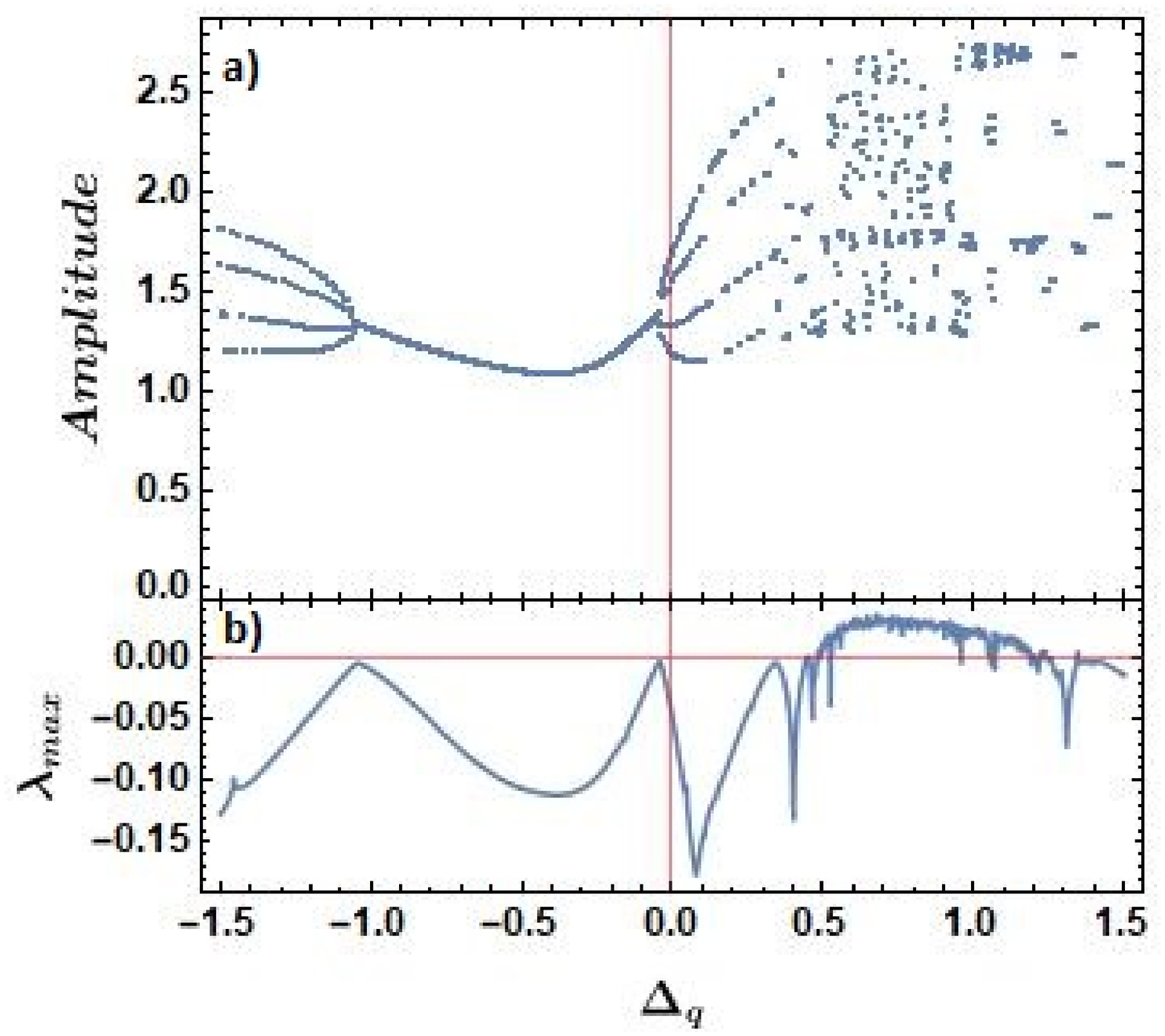}
    \caption{a) The Bifurcation Diagram for the system as we change the value of $\Delta_{q}$ b) Maximum Lyapunov Exponent for changing values of $\Delta_{q}$.  Here $\Delta=-0.75$, $\omega_{m}=1$, $\kappa=1$, $\kappa_{m}=0.001$, $\kappa_{q}=1$, $P=1.4$, $P_{p}=0.5$, $J=0.2$ and $\phi=0$}
    \label{Fig4}
\end{figure}
We can similarly change the value of power given to the qubit ($P$) given in \ref{Fig5}. The bifurcation diagram for this is quite erratic but we can again see chaotic motion in the system and period doubling bifurcation.\\
\begin{figure}[ht!]
    \includegraphics[width=0.45\textwidth]{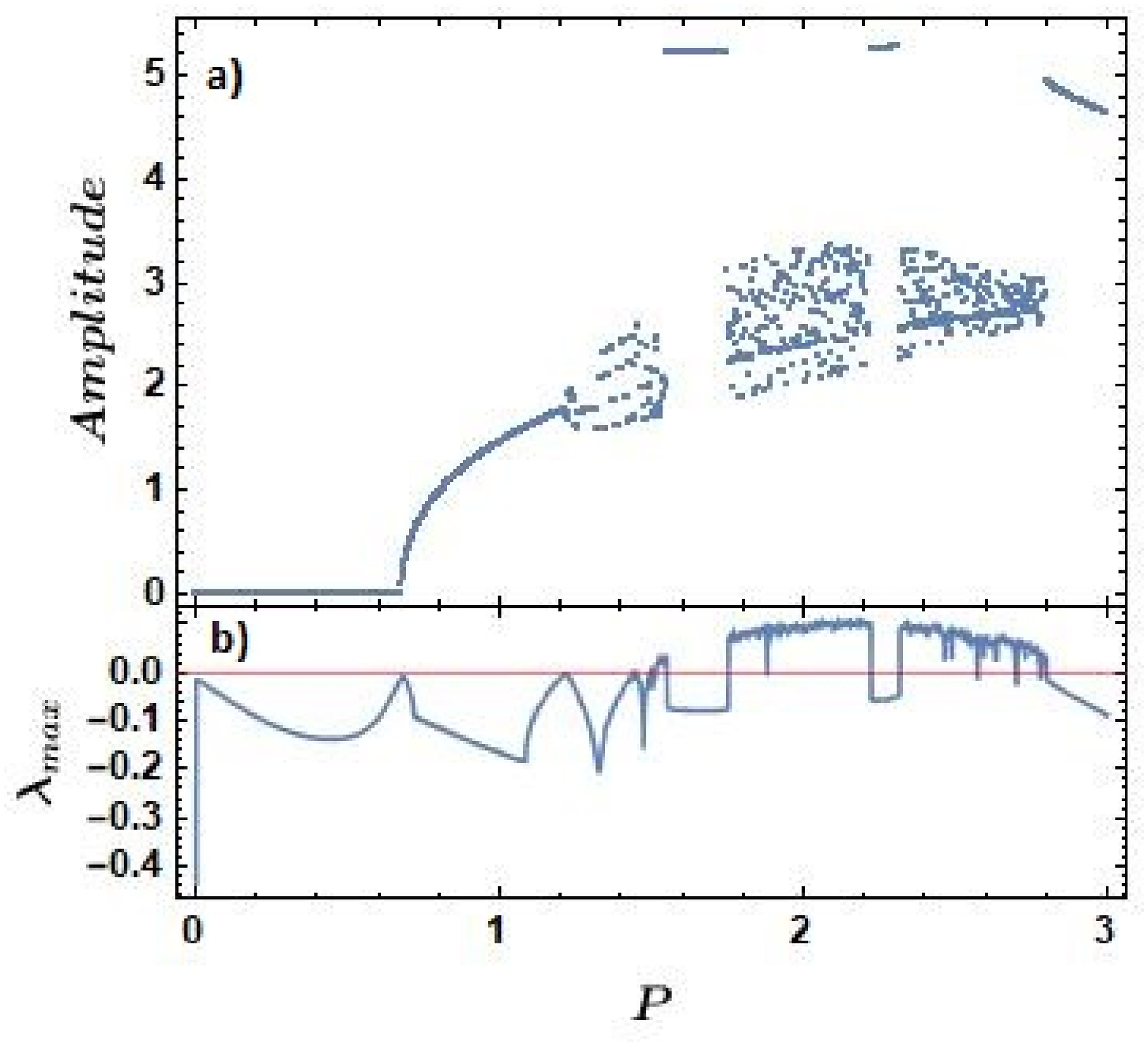}
    \caption{a) The Bifurcation Diagram for the system as we change the value of $P$ b) Maximum Lyapunov Exponent for changing values of $P_{p}$. Here $\Delta=-0.65$, $\Delta_{q}=0.5$, $\omega_{m}=1$, $\kappa=1$, $\kappa_{m}=0.001$, $\kappa_{q}=1$, $P_{p}=0.5$, $J=0.32$ and $\phi=0$}
    \label{Fig5}
\end{figure}
\begin{figure}[ht!]
    \includegraphics[width=0.45\textwidth]{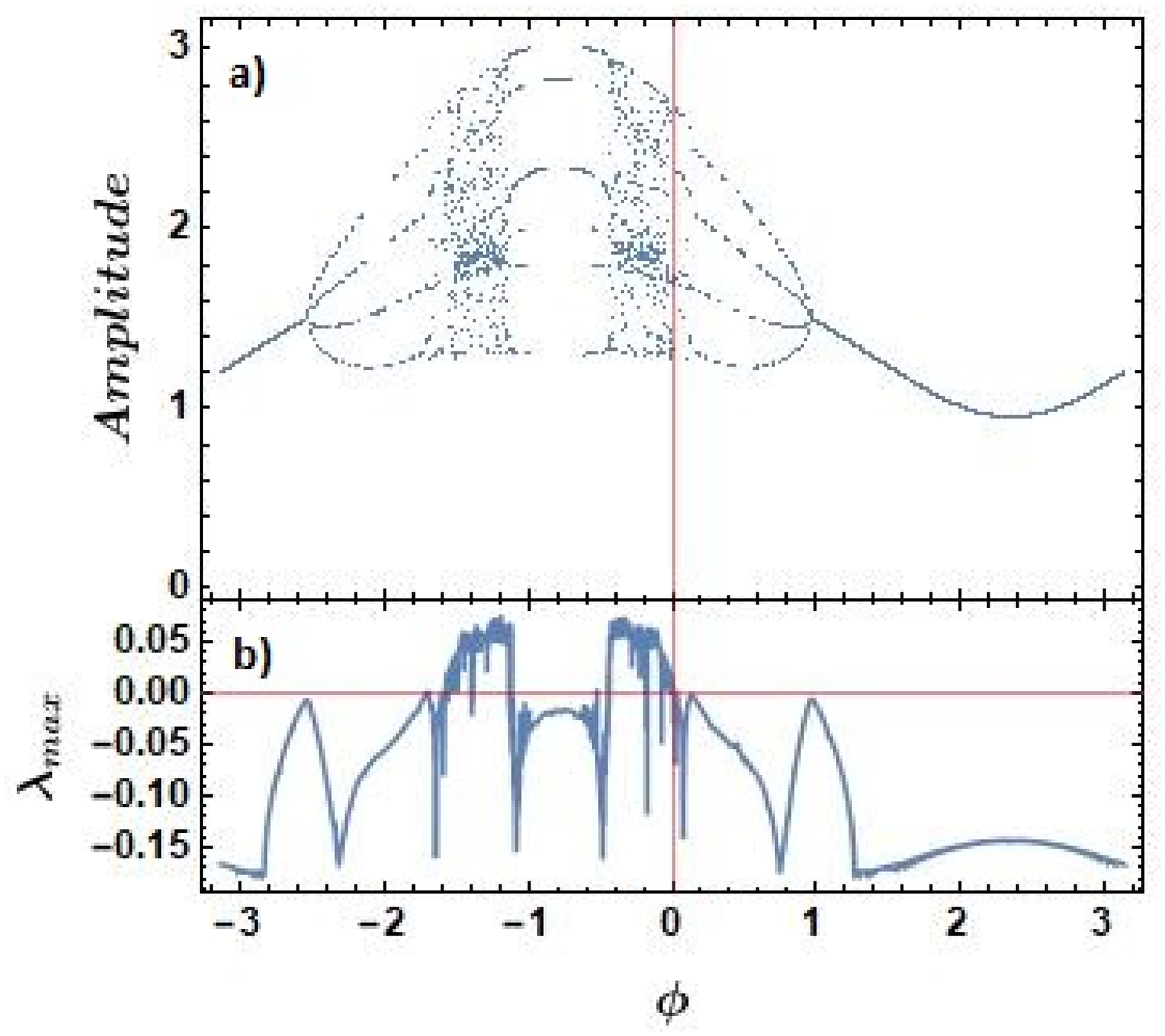}
    \caption{a) The Bifurcation Diagram for the system as we change the value of $\phi$ b) Maximum Lyapunov Exponent for changing values of $\phi$,. Here $\Delta=-0.65$, $\Delta_{q}=0.5$, $\omega_{m}=1$, $\kappa=1$, $\kappa_{m}=0.001$, $\kappa_{q}=1$, $P=1.4$, $J=0.2$ and $P_{p}=0.5$}
    \label{Fig6}
\end{figure}\\
However, the clearest example of chaotic nature can be seen while changing the value of phase difference ($\phi$) in \ref{Fig6}. This is experimentally easy to do since we only need to change the phase difference between the driving fields. We can see in Figure 6 that we can observe period-1, period-4, period-n and chaotic motion. It is also easily possible to tune in and out chaos here. Hence we can see how a qubit coupled optomechanical system can be used to generate chaos in an optomechanical system with ease.\\
Therefore in this work, we have shown various methods of generating Chaos in a qubit coupled optomechanical system. These methods make it easier than traditional methods to tune in and out of chaos using whichever setup is desired at the time. Producing these chaotic motions is also very easy since we can find chaotic motion using phase difference($\phi$), power($P$ and $P_{p}$), detuning($\Delta$ and $\Delta_{q}$) and even coupling ($J$). These can also be achieved experimentally since we can use the value of $w_{m}$ as desired in our experimental system. Using the value of $w_{m}=144.51 MHz$\cite{b.p19} we can get values of parameters which have been experimentally achieved before, by simply keeping ratio between $w_{m}$ and the desired parameter as found by us. Various different values can used as required by different setups. This coupled with the ability to achieve chaotic motion using various parameters allows for an easy setup to experimentally create chaotic qubits. Since the traditional setup of creating coupling a qubit with an optomechanical system involves using an LC circuit with a Josephson junction it is far easier to change the coupling parameter($J_{c}$) using different values of inductors and capacitors.\\
This allows us to create chaos with ease and further allows us to create chaotic motion in a qubit. This could be used for secret communication using qubits and quantum computers which would allow us to create more secure communication. This chaotic motion can also help us tune the optical cavity or qubit in and out of chaos far more easily since we only need to change the relative phase. Hence we can induce chaos in either the optical cavity or qubit without directly interfering with the system but changing the phase of the other system.

\acknowledgments


\begin{thebibliography}{0}

\bibitem{b.p1}
  \Name{Aspelmeyer M., Kippenberg T. J. \and Marquardt F.}
  \REVIEW{Rev. Mod. Phys.}{86}{2014}{1391}.
  
\bibitem{b.p2}
  \Name{Meystre P.}
  \REVIEW{Ann. Phys.}{525}{2013}{215}.

\bibitem{b.p3}
  \Name{Marquardt F. \and Girvin S. M.}
  \REVIEW{Physics 2}{40}{2009}.
  
\bibitem{b.p4}
  \Name{Kippenberg T. J. \and Vahala K. J.}
  \REVIEW{Science}{321}{2008}{1172}.

\bibitem{b.p5}
  \Name{Aspelmeyer M., Meystre P. \and Schwab K.}
  \REVIEW{Phys. Today}{65(7)}{2012}{2935}.
  
\bibitem{b.p6}
  \Name{Pepper B., Jeffrey E., Ghobadi R., Simon C. \and Bouwmeester D.}
  \REVIEW{Phys. Today}{65(7)}{2012}{2935}.
  
\bibitem{b.p7}
  \Name{Romero-Isart O.}
  \REVIEW{Phys. Rev.}{84}{2011}{052121}.
  
\bibitem{b.p8}
  \Name{Vanner M.R. {\it et al.}}
  \REVIEW{PNAS}{108(39)}{2011}{16182}.

\bibitem{b.p9}
  \Name{Kippenberg T. J., Rokhsari H., Carmon T., Scherer A. \and Vahala K. J.}
  \REVIEW{Phys. Rev. Lett.}{95}{2005}{033901}.
  
\bibitem{b.p10}
  \Name{Carmon T., Rokhsari H., Yang, L., Kippenberg T. J. \and Vahala K. J.}
  \REVIEW{Phys. Rev. Lett.}{94}{2005}{223902}.

\bibitem{b.p11}
  \Name{Marquardt F., Harris J. G. E. \and Girvin, S. M.}
  \REVIEW{Phys. Rev. Lett.}{96}{2006}{103901}.
  
\bibitem{b.p12}
  \Name{Metzger C. {\it et al.}}
  \REVIEW{Phys. Rev. Lett.}{101}{2008}{133903}.

\bibitem{b.p13}
  \Name{Zaitsev S., Pandey A. K., Shtempluck O. \and Buks, E. }
  \REVIEW{Phys. Rev. E}{84}{2011}{046605}.
  
\bibitem{b.p14}
  \Name{Carmon T., Rokhsari H., Yang L., Kippenberg T. J., \and Vahala K. J.}
  \REVIEW{Phys. Rev. Lett.}{94}{2005}{223902}.
  
\bibitem{b.p15}
  \Name{Carmon T., Cross M. C., \and Vahala K. J.}
  \REVIEW{Phys. Rev. Lett.}{98}{2007}{167203}.
  
\bibitem{b.p16}
  \Name{Navarro-Urrios D. {\it et al.}}
  \REVIEW{Nat Commun}{8}{2017}{14965}.

\bibitem{b.p17}
  \Name{Zhang D.-W., You C., \and Lü X.-Y.}
  \REVIEW{Phys. Rev. A}{101}{2020}{053851}.

\bibitem{b.p18}
  \Name{Yang N., Miranowicz A., Liu Y.-C., Xia  K. \and Nori F.}
  \REVIEW{Sci. Rep.}{9}{2019}{15874}.
  
\bibitem{b.p19}
  \Name{Lu X.-Y., Jing H., Ma J.-Y., \and Wu Y.}
  \REVIEW{Phys. Rev. Lett.}{114}{2015}{253601}.
  
\bibitem{b.p20}
  \Name{Zhang K., Chen W., Bhattacharya M., \and Meystre P.}
  \REVIEW{Phys. Rev. A}{81}{2010}{013802}.
  
\bibitem{b.p21}
  \Name{Wang M. {\it et al.}}
  \REVIEW{Sci. Rep.}{6}{2016}{22705}.

\bibitem{b.p22}
  \Name{Heinrich G., Ludwig M., Qian J., Kubala B., \and Marquardt F.}
  \REVIEW{Phys. Rev. Lett.}{107}{2011}{043603}.

\bibitem{b.p23}
  \Name{Xu X.-W., Lu X.-Y., Sun C.-P., \and Li Y.}
  \REVIEW{Phys. Rev. A}{ 92}{2015}{013852}.
  
\bibitem{b.p24}
  \Name{Regal C. A. \and Lehnert K. W.}
  \REVIEW{J. Phys.: Conf. Ser.}{264}{2011}{012025}.
  
\bibitem{b.p25}
  \Name{Taylor J. M., Sørensen A. S., Marcus C. M., \and Polzik E. S.}
  \REVIEW{Phys. Rev. Lett.}{107}{2011}{273601}.

\bibitem{b.p26}
  \Name{Brennecke F., Ritter S., Donner T. \and Esslinger T.}
  \REVIEW{Science}{322}{2008}{235}.
  
\bibitem{b.p27}
  \Name{Pirkkalainen, J. {\it et al.}}
  \REVIEW{Nat Commun}{6}{2015}{6981}.

\bibitem{b.p28}
  \Name{Wang H., Gu X., Liu  Y.-X., Miranowicz A., \and Nori F.}
  \REVIEW{Phys. Rev. A}{92}{2015}{033806}.
  
\bibitem{b.p29}
  \Name{Korppi M. {\it et al.}}
  \REVIEW{EPJ Web Conf.}{57}{2013}{03006}.

\bibitem{b.p30}
  \Name{Sivaprakasam S. \and Shore K. A.}
  \REVIEW{Opt. Lett.}{24}{1999}{466}.
  
\bibitem{b.p31}
  \Name{VanWiggeren G. D. \and Roy R.}
  \REVIEW{Science}{279}{1998}{1198}.
  
\bibitem{b.p32}
  \Name{Sciamanna M. \and Shore K. A.}
  \REVIEW{Nat. Photonics}{9}{2015}{151}.

\bibitem{b.p33}
  \Name{Redding B. {\it et al.}}
  \REVIEW{Proc. Natl. Acad. Sci.}{112}{2015}{1304}.
  
\bibitem{b.p34}
  \Name{Uchida A. {\it et al.}}
  \REVIEW{Nat. Photonics}{2}{2008}{728}.

\bibitem{b.p35}
  \Name{Law C. K.}
  \REVIEW{Phys. Rev. A}{51}{1995}{2537}.
  
\bibitem{b.p36}
  \Name{Farooq K. {\it et al.}}
  \REVIEW{Int. J. Mod. Phys. B}{33}{2019}{1950252}.

\bibitem{b.p37}
  \Name{Julsgaard B. \and Molmer K.}
  \REVIEW{Phys. Rev. A}{85}{2012}{013844}.
  
\bibitem{b.p38}
  \Name{Persico F. \and Vetri G.}
  \REVIEW{Phys. Rev. A}{12}{1975}{2083}.
  
\bibitem{b.p39}
  \Name{Marquardt F., Harris J. G. E. \and Girvin S. M.}
  \REVIEW{Phys. Rev. Lett.}{96}{2006}{103901}.

\bibitem{b.p40}
  \Name{Ludwig M., Kubala B. \and Marquardt F.}
  \REVIEW{New J. Phys.}{10}{2008}{095013}.

\bibitem{b.p41}
  \Name{Bakemeier L., Alvermann A. \and Fehske H.}
  \REVIEW{Phys. Rev. Lett.}{114}{2015}{013601}.
  
\bibitem{b.p42}
  \Name{Benettin G., Galgani L., Giorgilli A. \and Strelcyn J.-M.}
  \REVIEW{Meccanica}{15}{1980}{9}.

\bibitem{b.p43}
  \Name{Benettin G., Galgani L., Giorgilli A. \and Strelcyn J.-M.}
  \REVIEW{Meccanica}{15}{1980}{21}.
\end{thebibliography}
\end{document}